\documentclass[10pt,twocolumn,twoside,fleqn]{IEEEtran}
\ifCLASSINFOpdf
\else
\fi

\usepackage{graphicx}
\usepackage{amsmath,amsthm}
\usepackage{amssymb}
\usepackage{array}
\usepackage{color}
\usepackage[figuresright]{rotating}
\usepackage{subfigure}
\usepackage{graphics}
\usepackage[hidelinks]{hyperref}
\usepackage[normalem]{ulem}
\usepackage[para,online,flushleft]{threeparttable}
\usepackage{fontenc}
\usepackage[numbers,sort&compress]{natbib}

\newcommand{\tr}[1]{\textbf{tr} \left( #1 \right) }

\newcommand{\UV}[1]{\ensuremath{\mathbf{I}_{#1}}}
\newcommand{\AO}[1]{\ensuremath{\mathbf{1}_{#1}}}

\newcommand{\Lap}{\ensuremath{\mathbf{L}}}
\newcommand{\A}{\ensuremath{\mathbf{A}}}
\newcommand{\D}{\ensuremath{\mathbf{D}}}
\newcommand{\Prj}{\ensuremath{\mathbf{\Pi}}}
\newcommand{\T}{\ensuremath{\top}}

\newcommand{\Fr}{{\cal F}}

\newtheorem{theorem}{Theorem}[section]
\newtheorem{definition}{Definition}[section]
\newtheorem{lemma}[theorem]{Lemma}

\def\Nsf2{{N}_{\mbox{\scriptsize \rm{2SF}}}}

\newcommand{\spe}[1]{{#1}}

\begin{document}

\title{Scale-free Loopy Structure is Resistant to Noise in Consensus Dynamics in Complex Networks}
\author{Yuhao~Yi, Zhongzhi~Zhang, and Stacy~Patterson,~\IEEEmembership{Member,~IEEE}
\thanks{Yuhao Yi and Zhongzhi Zhang were supported by the National Natural Science
Foundation of China under Grant No. 11275049. Stacy Patterson was supported by NSF grants CNS-1527287 and CNS-1553340.}
\thanks{Yuhao Yi and Zhongzhi Zhang (corresponding author) are with the Shanghai Key Laboratory of Intelligent Information
Processing, School of Computer Science, Fudan University, Shanghai, 200433, China.
{\tt\small yhyi15@fudan.edu.cn}, {\tt\small zhangzz@fudan.edu.cn} }
\thanks{Stacy Patterson is with the Department of Computer Science, Rensselaer Polytechnic Institute, Troy, New York, 12180.
{\tt\small sep@cs.rpi.edu}
}
}

%



\IEEEtitleabstractindextext{%
\begin{abstract}
The vast majority  of real-world  networks are scale-free, loopy, and sparse, with a power-law degree distribution and a constant average degree. In this paper, we study first-order consensus dynamics in binary  scale-free networks, where vertices are subject to white noise. We focus on the coherence of networks characterized in terms of the $H_2$-norm, which quantifies how closely agents track the consensus value. We first provide a lower bound of coherence of a network in terms of its average degree, which is independent of the network order. We then study the coherence of some sparse, scale-free real-world networks, which approaches a constant. We also study numerically the coherence of  Barab\'asi-Albert networks and high-dimensional random Apollonian networks, which also converges to a constant when the networks grow.  Finally, based on the connection of coherence and the Kirchhoff index, we study analytically the coherence of two deterministically-growing sparse networks and obtain the exact expressions, which tend to small constants. Our results indicate that  the effect of noise on the consensus dynamics in power-law networks  is negligible. We argue that scale-free topology, together with loopy structure, is responsible for the strong robustness with respect to noisy consensus dynamics in power-law networks.
\end{abstract}

}

\maketitle

\IEEEdisplaynontitleabstractindextext

%
\IEEEpeerreviewmaketitle

\section{Introduction}

\IEEEPARstart{T}{he consensus} problem, as a fundamental problem in distributed computing~\cite{Ly97}, decision~\cite{Ts84} and control~\cite{SaFaMu07}, has been intensely studied in the context of networks of agents. It describes the process \spe{by} which a group of agents reach an agreement on certain quantities, for example, time, positions and attitudes of satellites, estimation of environmental quantities in a sensor network, and so on. \spe{Variations}  of the consensus problem can be observed in many science and engineering scenarios, including problems in clock synchronization~\cite{SuStErBrGh15}, load balancing~\cite{Cy89,MuGhSc98,DiFrMo99,AmFrJiVe15}, vehicle formation~\cite{FaMu04}, flocking~\cite{Sa06}, rendezvous~\cite{DiKy07}, human group dynamics~\cite{GiPa16}, distributed sensor networks~\cite{LiRu06,YuChWaYa09,ZhChLiYaGu13}, as well as distributed learning~\cite{AL14, MaBrMaSa16} and collaborative  inference~\cite{BiBlCa16} with streaming data.
In view of the wide range of  applications, consensus problems have attracted a great deal of recent interest~\cite{SaFaMu07,MoTa14,WuTaCaZh16}.

In real-world applications, agents are often subject to environmental disturbances.  For example, the motion of a group of vehicles is affected by many environmental factors, frictions, wind, slopes, to name a few, and these quantities can fluctuate within some range. In consensus problems with disturbances, agents never reach perfect equilibrium but fluctuate around the average of their current values. It is then ideal that the deviation for each agent from the average value is small, which is referred to as network coherence~\cite{PaBa10,PaBa11,BaJoMiPa12,PaBa14}. Due to its practical significance, consensus problems under environmental disturbances have attracted considerable attention~\cite{XiBoKi07,BaJoMiPa08,YoScLe10,RaWa11,YaBl13,YiZhLiCh15,HeZhChShCh16,YiZhShCh17}. In this paper, we  focus on a consensus protocol with Gaussian white noises added on the first-order derivatives of the vertex states, which is the most often discussed model in related \spe{literature}.


By convention, we use the mean steady state variance of the system to capture the average deviation of every vertex. This quantity is also called the \emph{first-order network coherence}, or \emph{coherence} for short in this paper.  By defining the output as the residual between the current states of vertices and their average, the coherence can be expressed by the $H_2$ norm of the system divided by the number of nodes. The $H_2$ norm of the considered system is given by the trace of the \spe{pseudoinverse} of the graph Laplacian matrix, which contains some critical information about the global topology of the network~\cite{PaBa11,BaJoMiPa12,PaBa14}. Note that the convergence rate in consensus \spe{problems} without noise is determined by the algebraic connectivity, which is the smallest none-zero eigenvalue of the graph Laplacian, while the coherence is related to the whole \spe{spectrum}.

Previous works have studied the network coherence in some particular network structures~\cite{PaBa10,PaBa11,BaJoMiPa12,PaBa14,XiBoKi07,BaJoMiPa08,YoScLe10,YiZhLiCh15}. Young \textit{et al.}~\cite{YoScLe10} gave closed-form solutions for the $H_2$ norm in paths, star graphs, directed and undirected cycles. Bamieh \textit{et al.}~\cite{PaBa11,PaBa14} studied the network coherence in some fractal trees and found \spe{a} connection between spectral dimensions and the first-order network coherence. They also give asymptotical results for coherence in tori and lattices~\cite{BaJoMiPa12}. To exploit the impact of  small-world and scale-free topology on the network coherence, we analytically determined the coherence in the small-world Farey network~\cite{ZhCo11,YiZhLiCh15} and the scale-free Koch network~\cite{YiZhShCh17}, with the latter consisting of short cycles \spe{of} only triangles.

However, the above-mentioned models cannot well mimic \spe{many real-world} networks, which are sparse and simultaneously display some remarkable properties~\cite{Ne03}. For example, various real-world networks exhibit the scale-free behavior~\cite{BaAl99} and small-world effect~\cite{WaSt98}. Scale-free behavior~\cite{BaAl99} means that the degree distribution $P(d)$ of many real-world networks has a power-law form as $P(d) \sim d^{-\gamma}$; while \spe{small-world} effect~\cite{WaSt98} implies that the average path length grows logarithmically with the number of vertices, or more slowly, and the average clustering coefficient tends to a constant larger than zero. In addition, \spe{many real-world} networks display a nontrivial pattern with many cycles/loops at different scales~\cite{RoKiBoBe05,KlSt06}, where a cycle/loop is  a sequence of different nodes  (except the starting node and ending node that are the same), with each pair of consecutive nodes in the sequence being adjacent to each other.  These striking structural patterns have great impact on other structural and dynamical properties of networks. For example, scale-free topology strongly affects structural characteristics (e.g., perfect matchings~\cite{ZhWu15} and minimum dominating sets~\cite{JiLiZh17}) and   dynamical processes (e.g., epidemic spreading~\cite{PaVe01}, game~\cite{SaSaPa08}, controllability~\cite{LiSlBa11}) on scale-free networks. 

As mentioned above, for first-order noisy consensus, it is desirable that the network coherence is as small as possible. Since \spe{many} real networks are sparse, \spe{natural questions arise: what} is the behavior of coherence for \spe{these} networks? What is the smallest \spe{possible} coherence for sparse \spe{networks} with given average degree? \spe{In addition}, realistic networks are \spe{often scale-free and small-world} with cycles at different scales. We then propose another interesting question: is \spe{there a} lower bound for coherence or its dominant scaling \spe{that} can be obtained in both real networks and popular models describing real networks with these prominent structural properties?

In this paper, we study the first-order coherence of noisy consensus on sparse networks with an average degree $\rho$, especially scale-free networks with cycles at distinct scales. First, we provide a lower bound and an upper bound for coherence of an arbitrary network, with the lower bound being a constant $1/(2\rho)$, independent of \spe{the} network order. Then, we consider the coherence of sparse real networks, which is also a constant but a little larger than $1/(2\rho)$. In addition, we address the  coherence of random, sparse scale-free \spe{network models}, including \spe{the} Barab{\'a}si-Albert (BA) network~\cite{BaAl99} and high dimensional random Apollonian networks (HDRAN)~\cite{ZhYaWa05,ZhRoFr06}, which is also constant. Finally, by making use of the connection between \spe{the} Kirchhoff index and the first-order network coherence, we study coherence on two exactly solvable, deterministic scale-free networks: one is the pseudofractal scale-free web~\cite{DoGoMe02}; the other is a new network proposed by the authors, which is called \spe{the} 4-clique motif scale-free network, hereafter, since it has a power-law distribution and consists of  4-vertex complete graphs. For both networks, we obtain the explicit expressions for their coherence, which also tends to constants as the networks grow. We show that the structural properties are responsible for the small coherence on these studied networks.

\section{Preliminaries}\label{coherence.sec}

In this section, we introduce some basic concepts in graph theory and electrical networks  and describe the consensus problems to be studied.

\subsection{Graph and Matrix Notation}

We consider a symmetric network system as an undirected graph $G$ \spe{consisting} of a pair $(V, E)$, where the vertex/node set $V=\{1,2,\cdots,N\}$ refers to $N$ nodes with dynamics, and the edge set $E$ contains $M$ unordered vertex pairs $\{i,j\}$, $i,j\in V$, representing links between vertices that can directly communicate. We use $|\cdot|$ to denote the cardinality of a set, thus $N=|V|$ and $M=|E|$. The adjacency matrix $\A$ is the matrix representation of graph $G$, which is defined as an $N \times N$ symmetric matrix with $a_{ij}=a_{ji}=1$ if the pair $\{i,j\}\in E$, and $a_{ij}=0$ otherwise. Let $\mathcal{N}_i$ be the set of neighbors for vertex $i$; then the degree of $i$ is $d_i=\sum_{j \in \mathcal{N}_i} a_{ij}=\sum_{j=1}^N a_{ij}$. The degree matrix $\D$ of graph $G$ is defined as a diagonal matrix with its $i$th diagonal entry  equal to $d_i$.   The Laplacian matrix of $G$ is  defined as $\Lap=\D-\A$. For a connected graph $G$, its Laplacian matrix $\Lap$ is positive semi-definite and has a unique zero eigenvalue corresponding to the eigenvector $\AO{N}$, which represents the $1\times N$ vector with all ones~\cite{Me98}. Thus all eigenvalues of $\Lap$ can be ordered as $0=\lambda_0<\lambda_1\leqslant\lambda_2\cdots\leqslant\lambda_{N-1}$ in a connected network. Moreover,  the sum of all non-zero eigenvalues is $2M$, that is   $\sum_{i=1}^{N-1}\lambda_i=2M$.

%

\subsection{Electrical \spe{Networks}}


An electrical network associated with graph $G$ is a network of resistances, where every edge in $G$ is replaced by a unit resistance. In the case without confusion, we also use $G$ to denote the electrical network  corresponding to graph $G$.  The \emph{resistance distance} between vertices $i$ and $j$ in $G$, denoted by $\Omega_{ij}$, is defined as the potential difference between them when a unit current is injected at $i$ and extracted from $j$. It has been proved that the resistance distance is a metric~\cite{KlRa93}. Then, the resistance distance between any pair of nodes is symmetric, that is, $\Omega_{ji}=\Omega_{ij}$ for two arbitrary vertices $i$ and $j$. It has been established~\cite{KlRa93} that  $\Omega_{ji}$ can be exactly represented in terms of the elements of the pseudoinverse  $\Lap^{\dagger}$ for $\Lap$:
\begin{align}
\label{Resij}
\Omega_{ji}=\Omega_{ij}=\Lap^{\dagger}_{ii}+\Lap^{\dagger}_{jj}-2\Lap^{\dagger}_{ij}\,.
\end{align}

The effective resistance of an electrical network has many interesting properties.
\begin{lemma}\label{Foster}
(Foster's Theorem~\cite{Fo49}) In an electrical network $G=(V,E)$,
\begin{align}
\sum_{\substack{i<j,\\(i,j)\in E}} \Omega_{ij} = N - 1.
\end{align}
\end{lemma}

\begin{lemma}\label{sumRule}
(Sum rule~\cite{Ch10}) For any two different vertices $i$ and $j$ in  an  electrical network $G=(V,E)$,
\begin{equation}
d_i\Omega_{ij}+\sum_{k\in \mathcal{N}_{i}}(\Omega_{ik}-\Omega_{jk})=2\,.
\end{equation}
\end{lemma}

For a network $G=(V,E)$, many graph invariants relevant to resistance distance have been defined.  Much studied examples include \spe{the} \emph{Kirchhoff index}~\cite{KlRa93},  \spe{the} \emph{multiplicative degree-Kirchhoff index}~\cite{ChZh07}, and \spe{the} \emph{additive degree-Kirchhoff index}~\cite{GuFeYu12}.  These three invariants are defined, respectively, by
\begin{equation}
\label{def:kir}
R(G)=\sum_{i,j\in V \atop{i<j}}\Omega_{ij}\,,
\end{equation}
\begin{equation}
\label{def:mulKir}
R^{\ast}(G)=\sum_{i,j\in V \atop{i<j}}d_i d_j \Omega_{ij}\,,
\end{equation}
and
\begin{equation}
\label{def:addKir}
R^{+}(G)=\sum_{i,j\in V \atop{i<j}}(d_i+d_j) \Omega_{ij}\,.
\end{equation}
Since $\Lap^{\dagger}\AO{N}=0$, from (\ref{Resij}) we obtain
\begin{align}
\label{def:KirB}
R(G)=\frac{1}{2}\sum_{i,j\in V}\Omega_{ij}=N\cdot \tr{\Lap^{\dagger}}=N\cdot \sum_{i=1}^{N-1} \frac{1}{\lambda_{i}}\,.
\end{align}

\subsection{First-Order Leader-Free Noisy Consensus Dynamics}

In the first-order consensus problem, the state of the system is given by a vector $x\in \mathbb{R}^N$, where the $i$-th entry $x_i \in \mathbb{R}$ denotes the state of vertex $i$. Let $x(t) \in \mathbb{R}^N$ denote the system state at time $t$. Each vertex utilizes only local information to adjust its state, and the states of  vertices are  subject to stochastic disturbances. The first-order dynamics in a stochastic consensus problem without any leader can be described by
\begin{align}\label{xdotv.eq}
\mathrm{d}x_i(t)&=-\sum_{j \in {{\mathcal{N}}_i}} a_{ij}[x_i(t)-x_j(t)]\mathrm{d}t+\mathrm{d}W_i(t)\,,
\end{align}
in which  $W_i(t)$ is a Wiener process. Let $W(t)$ be the vector of $N$ uncorrelated Wiener processes; then, we can write~\eqref{xdotv.eq} in matrix form as
\begin{equation}\label{xdot.eq}
\mathrm{d}x(t)=-\Lap x(t)\mathrm{d}t+\mathrm{d}W(t)\,.
\end{equation}

It is known that when the agents are subject to external \spe{disturbances}, the vertex states fluctuate around the average of the states of all vertices. We characterize the variance of these fluctuations by the concept of network coherence~\cite{PaBa11,BaJoMiPa12,PaBa14}.

\begin{definition}
The \emph{first-order network coherence} without leaders is defined as the mean steady-state variance of the deviation from the average of the current vertex states:
\begin{equation}\label{cohDefIn}
H_\mathrm{FO} :=  \lim_{t \rightarrow \infty}  \frac{1}{N}  \sum_{i=1}^{N}\mathbf{var}\left \{x_{i}(t) - \frac{1}{N}\sum_{j=1}^{N} x_{j}(t)\right \} \,.
\end{equation}
\end{definition}

The output of the system is given by
\begin{equation}
y(t) = \Prj \,x(t),
\label{output.eq}
\end{equation}
where $\Prj$ is the projection operator defined as ${\Prj:=\UV{N}-\frac{1}{N}\AO{N} \AO{N}^\T}$, with $\UV{N}$ being the identity matrix of order $N$. 
$H_\text{FO}$ is related to an $H_2$ norm~\cite{YoScLe10} formulated by~\eqref{xdot.eq} and~\eqref{output.eq}:
\begin{small}
\begin{equation}\label{focoh1.eq}
H_\text{FO} =\frac{(H_2)^2}{N}=\frac{1}{N} \tr{\int_0^{\infty} \mathrm{e}^{-\Lap^\T t}\Prj^\T\Prj \mathrm{e}^{-\Lap t}\mathrm dt}=\frac{1}{2N}\tr{\Lap^{\dagger}}\,.
\end{equation}
\end{small}
According to~\eqref{def:KirB},  $H_\text{FO}$ depends on the $N-1$ nonzero eigenvalues of Laplacian matrix $\Lap$, \spe{and similarly,}  the Kirchhoff index $R(G)$  for  graph $G$:
\begin{equation}
\label{FOCohRes}
H_{\mathrm{FO}}(G)=\frac{1}{2 N} \sum_{i=1}^{N-1} \frac{1}{\lambda_{i}}=\frac{R(G)}{2N^2}\,.
\end{equation}

\subsection{Related Work}


The first-order network coherence $H_{\mathrm{FO}}$ and its scaling behavior in different networks have been extensively studied. Table~\ref{Scaling} lists the asymptotic scalings for $H_{\mathrm{FO}}$ in some networks previously studied in the literature.

\begin{table}
\caption{Scalings of average  graph  distance $\mu$  and  coherence  $H_{\mathrm{FO}}$  for  some typical network structures.}\label{Scaling}
\resizebox{\columnwidth}{!}{%

\begin{tabular}{|c|c|c|}
\hline
\raisebox{-0.5ex}{Network Structure} & \raisebox{-0.5ex}{ $\mu$ } & \raisebox{-0.5ex}{$H_{\rm{FO}}$} \\
		\hline
		\hline
		\raisebox{-0.5ex}{path~\cite{YoScLe10}} & \raisebox{-0.5ex}{$N$} & \raisebox{-0.5ex}{$N$}  \\[0.5ex]
		\hline
		\raisebox{-0.5ex}{1-dimensional torus~\cite{BaJoMiPa12, YoScLe10}} & \raisebox{-0.5ex}{$N$} & \raisebox{-0.5ex}{$N$} \\[0.5ex]
		\hline
		\raisebox{-0.5ex}{1-dimensional Cayley graph~\cite{LoZa10} } & \raisebox{-0.5ex}{$N$} & \raisebox{-0.5ex}{$N$} \\[0.5ex]
		\hline
		\raisebox{-0.5ex}{regular ring lattice~\cite{YiZhLiCh15}} & \raisebox{-0.5ex}{$N$} & \raisebox{-0.5ex}{$N$} \\[0.5ex]
		\hline
		\raisebox{-0.5ex}{ Vicsek fractal~\cite{PaBa14}} & \raisebox{-0.5ex}{$N^{\log{3}/\log{5}}$} & \raisebox{-0.5ex}{$N^{\log{3}/\log{5}}$} \\[0.5ex]
		\hline
		\raisebox{-0.5ex}{T-fractal~\cite{PaBa14}} & \raisebox{-0.5ex}{$N^{\log{2}/\log{3}}$} & \raisebox{-0.5ex}{$N^{\log{2}/\log{3}}$} \\[0.5ex]
		\hline
		\raisebox{-0.5ex}{Peano Basin fractal~\cite{PaBa14}} & \raisebox{-0.5ex}{$N^{1/2}$} & \raisebox{-0.5ex}{$N^{1/2}$} \\[0.5ex]
		\hline
		\raisebox{-0.5ex}{2-dimensional torus~\cite{BaJoMiPa12} } & \raisebox{-0.5ex}{$N^{1/2}$} & \raisebox{-0.5ex}{$\log{N}$} \\[0.5ex]
		\hline
		\raisebox{-0.5ex}{2-dimensional Cayley graph~\cite{LoZa10} } & \raisebox{-0.5ex}{$N^{1/2}$} & \raisebox{-0.5ex}{$\log{N}$} \\[0.5ex]
		\hline
		\raisebox{-0.5ex}{Farey graph~\cite{YiZhLiCh15}} & \raisebox{-0.5ex}{$\log{N}$} & \raisebox{-0.5ex}{$\log{N}$} \\[0.5ex]
		\hline
		\raisebox{-0.5ex}{Koch graph~\cite{YiZhShCh17}} & \raisebox{-0.5ex}{$\log{N}$} & \raisebox{-0.5ex}{$\log{N}$} \\[0.5ex]
		\hline
		\raisebox{-0.5ex}{$d$-dimensional torus ($d\geqslant 3$)~\cite{BaJoMiPa12} } & \raisebox{-0.5ex}{$N^{1/d}$} & \raisebox{-0.5ex}{$1$} \\[0.5ex]
		\hline
		\raisebox{-0.5ex}{$d$-dimensional  Cayley graph ($d\geqslant 3$)~\cite{LoZa10} } & \raisebox{-0.5ex}{$N^{1/d}$} & \raisebox{-0.5ex}{$1$} \\[0.5ex]
		\hline
		\raisebox{-0.5ex}{star graph~\cite{YoScLe10}} & \raisebox{-0.5ex}{$1$} & \raisebox{-0.5ex}{$1$} \\[0.5ex]
		\hline
		\raisebox{-0.5ex}{complete graph~\cite{YoScLe10}} & \raisebox{-0.5ex}{$1$} & \raisebox{-0.5ex}{$N^{-1}$} \\[0.5ex]
		\hline
	\end{tabular}
	}

\end{table}

From Table~\ref{Scaling}, we can observe that the leading behavior for $H_{\mathrm{FO}}$ in different networks is rich. For a network with $N$ vertices, $H_{\mathrm{FO}}$ can behave linearly, sub-linearly, logarithmically, inversely with $N$, or independently of $N$. For example, in the path graph~\cite{YoScLe10}, 1-dimensional torus~\cite{BaJoMiPa12, YoScLe10},  1-dimensional Cayley graph~\cite{LoZa10}, and regular ring lattice~\cite{YiZhLiCh15}, $H_{\mathrm{FO}}\sim N$; in some fractal tree-like graphs~\cite{PaBa14} including Vicsek fractal, T fractal, and Peano Basin fractal, $H_{\mathrm{FO}}$ grows sub-linearly with $N$ as $ H_{\mathrm{FO}} \sim  N^\theta$ with $ 0 <\theta <1$; in 2-dimensional torus~\cite{BaJoMiPa12}, 2-dimensional Cayley graph~\cite{LoZa10}, Farey graph~\cite{YiZhLiCh15}, and Koch graph~\cite{YiZhShCh17},  $ H_{\mathrm{FO}} \sim \ln N$; in the complete graph~\cite{YoScLe10}, $ H_{\mathrm{FO}} \sim 1/N$; while in $d$-dimensional torus ($d\geqslant 3$)~\cite{BaJoMiPa12}, $d$-dimensional Cayley graph ($d\geqslant 3$)~\cite{LoZa10}, as well as star graph~\cite{YoScLe10},  $H_{\mathrm{FO}}$ is a constant, irrespective of $N$.

It can be proved~\cite{YoScLe10} that among all $N$-vertex graphs, the first-order network coherence $H_{\mathrm{FO}}$ is minimized uniquely  in the complete graph $K_N$, with $H_{\mathrm{FO}}(K_N)=\frac{N-1}{2N^2}$. When $N \rightarrow \infty$, $H_{\mathrm{FO}}(K_N) \rightarrow 0$. In this sense, the complete graph has the optimal structure that has the best performance for noisy consensus dynamics.   However, complete graphs are dense, with  the degree of each vertex being $N-1$. Extensive empirical work indicates that real-world networks are often sparse, having a small constant average degree~\cite{Ne03}. Moreover, most realistic networks are scale-free~\cite{BaAl99} and small-world~\cite{WaSt98}. Table~\ref{Scaling} shows that the first-order network coherence $H_{\mathrm{FO}}$ depends on the structure of networks.  Then, interesting  questions arise naturally: What is the minimum scaling of $H_{\mathrm{FO}}$ for sparse networks? Is this minimal scaling  achieved in real scale-free  networks?

In the sequel, we first provide a lower bound $1/(2\rho)$ for  the first-order network coherence $H_{\mathrm{FO}}$ for all networks with average degree $\rho$. Then we study   the first-order network coherence $H_{\mathrm{FO}}$ for some real scale-free networks and show that $H_{\mathrm{FO}}$ is constant. We also study   the first-order network coherence $H_{\mathrm{FO}}$ on two random scale-free networks, BA  network~\cite{BaAl99} and HDRAN~\cite{ZhYaWa05,ZhRoFr06}, which converges to a constant. Finally, \spe{we} study analytically $H_{\mathrm{FO}}$ \spe{in} two deterministic scale-free networks:  pseudofractal scale-free web~\cite{DoGoMe02} and the 4-clique motif scale-free network, the  first-order network coherence for which also tends to a constant. Thus,  scale-free topology is advantageous to noisy consensus dynamics.

\section{Lower and upper bounds for  first-order network coherence}


In this section, we provide   a lower bound and an upper bound for first-order network coherence $H_{\mathrm{FO}}$ in  an arbitrary  graph. We first introduce a lemma~\cite{Me94}.
\begin{lemma} \label{isComplete}
Let  $G$ be an $N$-vertex graph. Then $\lambda_1 = \lambda_2 = \cdots = \lambda_{N-1}$ if and only if $G$ is a complete graph.
\end{lemma}

\begin{theorem}\label{TheoFO}
For a graph $G$  with $N$ vertices,  $M$ edges,  and average degree $\rho=\frac{2M}{N}$,  the first-order network coherence $H_{\mathrm{FO}} \geq \frac{N}{4M}-\frac{1}{2M}+\frac{1}{4MN}$, with equality  if and only if $G$ is the complete graph; \spe{further,} $H_{\mathrm{FO}} \geq \frac{1}{2\rho}$ when $N$ is large.
\end{theorem}
\begin{IEEEproof}
Applying the Cauchy-Schwarz inequality to~\eqref{FOCohRes} yields
\begin{align}
H_{\textrm{FO}}
=&\frac{1}{2N}\sum_{i=1}^{N-1}\frac{1}{\lambda_i}=\frac{1}{2N}\sum_{i=1}^{N-1}\frac{1}{\lambda_i}\sum_{j=1}^{N-1}\frac{\lambda_j}{2M}\nonumber\\
\geqslant & \frac{1}{4MN}\left(\sum_{i=1}^{N-1}\sqrt{\frac{1}{\lambda_i} \cdot \lambda_i}\right)^2=\frac{\left(N-1\right)^2}{4MN}\nonumber\\
=&\frac{N}{4M}-\frac{1}{2M}+\frac{1}{4MN}\,.
\end{align}
By Lemma~\ref{isComplete}, $H_{\mathrm{FO}}= \frac{N}{4M}-\frac{1}{2M}+\frac{1}{4MN}$  if and only if $G$ is the complete graph.

Since   $\rho=\frac{2M}{N}$ and $2M=\rho N\geqslant 2(N-1)$,
\begin{align}
\lim_{N\to \infty}H_{\mathrm{FO}} \geqslant \frac{1}{2\rho}\,.
\end{align}
This completes the proof.
\end{IEEEproof}

In addition to the lower bound, we  also provide an upper bound for  the  first-order network coherence $H_{\mathrm{FO}}$ of a graph $G$ in terms of its average \spe{graph} distance (also called  average path length) $\mu$.
\begin{theorem}
For a graph $G$  with $N$ vertices   and average \spe{graph} distance  $\mu$,  the first-order network coherence ${H_{\mathrm{FO}} \leqslant \frac{N-1}{4N}\mu}$, with the equality  if and only if $G$ is a tree.  When $N$ is large,  $H_{\mathrm{FO}} \leqslant \frac{\mu}{4}$.
\end{theorem}
\begin{IEEEproof}
Note that in any graph,  the effective resistance $\Omega_{ij}$ between a pair of  vertices $i$ and $j$  is less than or equal to their shortest path length $\delta_{ij}$ \cite{KlRa93}. Then,
\begin{equation}
H_{\mathrm{FO}} =\frac{R(G)}{2N^2} \leqslant  \frac{\mu\frac{N(N-1)}{2}}{2N^2}=\frac{N-1}{4N} \mu\,.
\end{equation}
When $G$ is a tree, $\Omega_{ij}=\delta_{ij}$ \cite{KlRa93}, and thus the equality holds.

For large $N$,
\begin{equation}
\lim_{N\to \infty}H_{\mathrm{FO}}  \leqslant \frac{\mu}{4}\,.
\end{equation}
This completes the proof.
\end{IEEEproof}

For a network  $G$, it is highly desirable that its first-order network coherence is small. Theorem~\ref{TheoFO} indicates that for large sparse graphs with small constant $\rho$,   $\frac{1}{2\rho}$ is the smallest value we can obtain for the  first-order network coherence $H_{\mathrm{FO}}$.  A network is \spe{said} to be \emph{optimal} if its   first-order network coherence is $\frac{1}{2\rho}$.  We refer to a large sparse network as \emph{almost optimal} if its first-order coherence is a constant.

Table~\ref{Scaling} shows that for \spe{the} $d$-dimensional torus ($d\geqslant 3$), \spe{the} $d$-dimensional Cayley graph ($d\geqslant 3$), and the star graph, the  first-order network coherence  is a constant.  In fact,  the star graph $S_N$ has the smallest $H_{\text{FO}}$ among all trees with $N$ vertices. $H_{\text{FO}}$ for $S_N$ does not grow with $N$ but converges to a constant. For  $d$-dimensional tori and $d$-Cayley graphs ($d\geqslant 3$), their $H_{\text{FO}}$ is also a constant, but their average distance grows with $N$.

As  mentioned above, most real-world networks are sparse and have a  power-law degree distribution. However, the behavior of  first-order network coherence for realistic networks has not been studied thus far.  In what follows, we will study the first-order network coherence for some real-life scale-free networks, \spe{as well as} random and deterministic  \spe{network models}. We will show that for all \spe{of} these considered sparse networks, their first-order network coherence does not \spe{grow} with the network size but converges to small constants.  Thus, scale-free networks have almost optimal network coherence.

\section{Network Coherence for Realistic  Networks}

In this section, we evaluate the coherence of some real-world networks that have power-law degree distributions. We use a large collection of networks of different orders that are chosen from different domains.

In Table \ref{Stat} we report the network coherence for some real-world,  scale-free undirected networks. All data are taken from the Koblenz Network Collection~\cite{Ku13}. The considered real networks are \spe{representative}, including social networks, information networks, technological networks, and metabolic networks. The networks are shown in increasing \spe{order} of the number of vertices. The smallest network has approximately $3\times 10^2$ vertices while the largest network has about $6\times 10^5$ vertices.

\begin{table}
\caption{Basic statistics and first-order coherence of some real-world networks.}\label{Stat}
\begin{tablenotes}
For each network, we  indicate  the number of nodes $N$, the number of edges $M$, power-law exponent $\gamma$, the lower bound for coherence $\frac{1}{2\rho}$, the  coherence for the largest connected component $H_\mathrm{FO}$, and the upper bound for coherence  $\frac{\mu}{4}$.
	\end{tablenotes}
\resizebox{\columnwidth}{!}{%
\begin{tabular}{|c|c|c|c|c|c|c|}
\hline
\raisebox{-0.5ex}{Network} & \raisebox{-0.5ex}{$N$} & \raisebox{-0.5ex}{$M$} & \raisebox{-0.5ex}{$\gamma$} & $\frac{1}{2\rho}$ & \raisebox{-0.5ex}{$H_{\textrm{FO}}$} & \raisebox{-0.5ex}{$\frac{\mu}{4}$} \\
		\hline
		\hline
		\raisebox{-0.5ex}{Zachary karate club} & \raisebox{-0.5ex}{34} & \raisebox{-0.5ex}{78} & \raisebox{-0.5ex}{2.161} & \raisebox{-0.5ex}{0.109} & \raisebox{-0.5ex}{0.203} & \raisebox{-0.5ex}{0.602} \\[0.5ex]
		\hline
		\raisebox{-0.5ex}{David Copperfield} & \raisebox{-0.5ex}{112} & \raisebox{-0.5ex}{425} & \raisebox{-0.5ex}{3.621} & \raisebox{-0.5ex}{0.066} & \raisebox{-0.5ex}{0.151}  & \raisebox{-0.5ex}{0.634} \\[0.5ex]
		\hline
		\raisebox{-0.5ex}{Hypertext 2009} & \raisebox{-0.5ex}{113} & \raisebox{-0.5ex}{2,196} & \raisebox{-0.5ex}{1.284} & \raisebox{-0.5ex}{0.013} & \raisebox{-0.5ex}{0.021} & \raisebox{-0.5ex}{0.414} \\[0.5ex]
		\hline
		\raisebox{-0.5ex}{Jazz musicians} & \raisebox{-0.5ex}{198} & \raisebox{-0.5ex}{2,742} & \raisebox{-0.5ex}{5.271} & \raisebox{-0.5ex}{0.018} & \raisebox{-0.5ex}{0.051} & \raisebox{-0.5ex}{0.559} \\[0.5ex]
		\hline
		\raisebox{-0.5ex}{PDZBase} & \raisebox{-0.5ex}{212} & \raisebox{-0.5ex}{242} & \raisebox{-0.5ex}{3.034} &\raisebox{-0.5ex}{0.109} & \raisebox{-0.5ex}{0.707} & \raisebox{-0.5ex}{1.332} \\[0.5ex]
		\hline
		\raisebox{-0.5ex}{Haggle} & \raisebox{-0.5ex}{274} & \raisebox{-0.5ex}{2,124} & \raisebox{-0.5ex}{1.673} & \raisebox{-0.5ex}{0.219} & \raisebox{-0.5ex}{0.236} & \raisebox{-0.5ex}{0.606} \\[0.5ex]
		\hline
		\raisebox{-0.5ex}{Caenorhabditis elegans} & \raisebox{-0.5ex}{453} & \raisebox{-0.5ex}{2,025} & \raisebox{-0.5ex}{1.566} &\raisebox{-0.5ex}{0.056} & \raisebox{-0.5ex}{0.135} & \raisebox{-0.5ex}{0.666} \\[0.5ex]
		\hline
		\raisebox{-0.5ex}{U. Rovira i Virgili} & \raisebox{-0.5ex}{1,133} & \raisebox{-0.5ex}{5,451} & \raisebox{-0.5ex}{1.561} &\raisebox{-0.5ex}{0.052}  & \raisebox{-0.5ex}{0.170} & \raisebox{-0.5ex}{0.902}  \\[0.5ex]
		\hline
		\raisebox{-0.5ex}{Hamsterster friendships} & \raisebox{-0.5ex}{1,858} & \raisebox{-0.5ex}{12,534} & \raisebox{-0.5ex}{2.461} &\raisebox{-0.5ex}{0.037}  & \raisebox{-0.5ex}{0.176} & \raisebox{-0.5ex}{0.863}  \\[0.5ex]
		\hline
		\raisebox{-0.5ex}{Protein} & \raisebox{-0.5ex}{1,870} & \raisebox{-0.5ex}{2,203} & \raisebox{-0.5ex}{2.879} & \raisebox{-0.5ex}{0.212} & \raisebox{-0.5ex}{0.730} & \raisebox{-0.5ex}{1.703} \\[0.5ex]
		\hline
		\raisebox{-0.5ex}{Hamster  full} & \raisebox{-0.5ex}{2,426} & \raisebox{-0.5ex}{16,631}& \raisebox{-0.5ex}{2.421} & \raisebox{-0.5ex}{0.037} & \raisebox{-0.5ex}{0.142} & \raisebox{-0.5ex}{0.897} \\[0.5ex]
		\hline
		\raisebox{-0.5ex}{Facebook (NIPS)} & \raisebox{-0.5ex}{2,888} & \raisebox{-0.5ex}{2,981} & \raisebox{-0.5ex}{4.521} & \raisebox{-0.5ex}{0.242} & \raisebox{-0.5ex}{0.675} & \raisebox{-0.5ex}{0.967} \\[0.5ex]
		\hline
		\raisebox{-0.5ex}{Human protein (Vidal)} & \raisebox{-0.5ex}{3,133} & \raisebox{-0.5ex}{6,149} & \raisebox{-0.5ex}{2.132} & \raisebox{-0.5ex}{0.127} & \raisebox{-0.5ex}{0.388} & \raisebox{-0.5ex}{1.210} \\[0.5ex]
		\hline
		\raisebox{-0.5ex}{Reactome} & \raisebox{-0.5ex}{6,327} & \raisebox{-0.5ex}{146,160} & \raisebox{-0.5ex}{1.363} & \raisebox{-0.5ex}{0.011} & \raisebox{-0.5ex}{0.138} & \raisebox{-0.5ex}{1.053} \\[0.5ex]
		\hline
		\raisebox{-0.5ex}{Route views} & \raisebox{-0.5ex}{6,474} & \raisebox{-0.5ex}{12,572} & \raisebox{-0.5ex}{2.462}& \raisebox{-0.5ex}{0.129} & \raisebox{-0.5ex}{0.365} & \raisebox{-0.5ex}{0.926} \\[0.5ex]
		\hline
		\raisebox{-0.5ex}{Pretty Good Privacy} & \raisebox{-0.5ex}{10,680} & \raisebox{-0.5ex}{24,316} & \raisebox{-0.5ex}{4.261} & \raisebox{-0.5ex}{0.110} & \raisebox{-0.5ex}{0.721}& \raisebox{-0.5ex}{1.871} \\[0.5ex]
		\hline
		\raisebox{-0.5ex}{arXiv astro-ph} & \raisebox{-0.5ex}{18,771} & \raisebox{-0.5ex}{198,050} & \raisebox{-0.5ex}{2.861} & \raisebox{-0.5ex}{0.024} & \raisebox{-0.5ex}{0.128} & \raisebox{-0.5ex}{1.049} \\[0.5ex]
		\hline
		\raisebox{-0.5ex}{CAIDA} & \raisebox{-0.5ex}{26,475} & \raisebox{-0.5ex}{53,381} & \raisebox{-0.5ex}{2.509} & \raisebox{-0.5ex}{0.124} & \raisebox{-0.5ex}{0.361} & \raisebox{-0.5ex}{0.969}  \\[0.5ex]
		\hline
		\raisebox{-0.5ex}{Internet topology} & \raisebox{-0.5ex}{34,761} & \raisebox{-0.5ex}{107,720} & \raisebox{-0.5ex}{2.233} & \raisebox{-0.5ex}{0.081} & \raisebox{-0.5ex}{0.319} & \raisebox{-0.5ex}{1.229} \\[0.5ex]
		\hline
		\raisebox{-0.5ex}{Brightkite} & \raisebox{-0.5ex}{58,228} & \raisebox{-0.5ex}{214,078} & \raisebox{-0.5ex}{2.481} &\raisebox{-0.5ex}{0.068} & \raisebox{-0.5ex}{0.359} & \raisebox{-0.5ex}{0.942} \\[0.5ex]
		\hline
				

	\end{tabular}
	}
\end{table}

Table~\ref{Stat} shows that for real scale-free networks with $2 < \gamma \leqslant 3$, their coherence $H_{\text{FO}}$ is generally small. For all networks, $H_{\text{FO}}$ lies between its lower bound  $1/(2\rho)$  and upper bound  $\mu/4$. Moreover, for each network, its $H_{\text{FO}}$ is significantly smaller than $\mu/4$; for most networks, $H_{\text{FO}}$ is closer to $1/(2\rho)$ than $\mu/4$. In fact, as observed for many other properties (e.g., clustering coefficient~\cite{WaSt98}), the coherence $H_{\text{FO}}$ of real networks does not increase with the number of vertices $N$; instead, it seems to be independent of $N$ and tends to small constants.

\section{Network Coherence in Random Scale-free Model Networks}

In this section, we study the coherence for two  typical stochastic scale-free model networks, that is, Barab\'{a}si-Albert (BA) networks~\cite{BaAl99} and  high dimensional random Apollonian networks (HDRAN)~\cite{ZhYaWa05,ZhRoFr06}. The main reason for studying these two networks is that they capture the generating mechanisms for some real scale-free networks. 

\subsection{Coherence for Barab\'{a}si-Albert Networks}

The BA network model~\cite{BaAl99} is the most well known random scale-free model. The algorithm of the BA model is as follows. Starting from a small connected graph, at each time step, we create a new vertex and connect it to $m$ different vertices that are already present in the network.  The probability that the new vertex connects to an old vertex $i$ is proportional to the degree of  $i$. We repeat the growth and  preferential attachment procedure, until the network grows to the ideal size  $N$. When $N$ is large, the average degree of  BA networks approaches $2m$. The degree distribution of BA networks exhibits a   power-law form, with the power exponent $\gamma$ being $3$, regardless of $m$.   BA networks  are small-world, with their average path length  increasing approximately logarithmically with $N$. In addition, there are many cycles with different lengths   in BA networks~\cite{BiCa03}.

\begin{figure}
  \centering\includegraphics[width=1.0\linewidth]{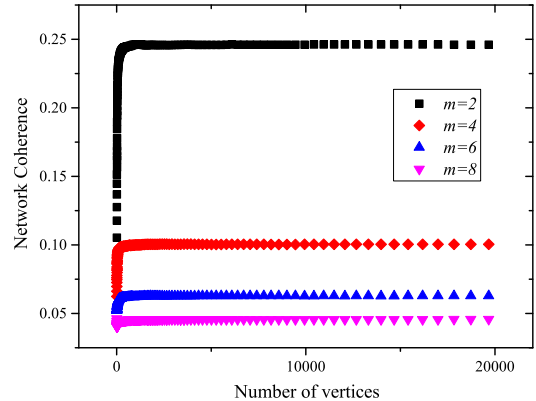}
  \caption{Simulation results for  the coherence of BA networks with different $m$. For all networks,  the construction starts from the $8$-vertex complete graph.}\label{BASim.fig}
\end{figure}

According to the generating algorithms,  we create different networks with various number of vertices. For all of these generated networks, we calculate their coherence based on \eqref{focoh1.eq}. Figure \ref{BASim.fig} shows the coherence of BA networks with various $m=2$, $4$, $6$,  $8$. From the figure, we observe that the coherence of these networks does not grow with the network size $N$ but converges to an $m$-dependent  constant: the larger the parameter $m$, the smaller the network coherence.

\subsection{Coherence for High Dimensional Random Apollonian Networks}

We continue to study the coherence for $d$-dimensional ($d\geq 2$) random Apollonian networks. We will see that the behavior of  coherence  for HDRAN is similar to that observed in BA networks. The   $d$-dimensional ($d\geq 2$) random Apollonian networks are constructed as follows. We first generate a $d+2$-vertex complete graph, or $d+2$-clique. We say that a $d+1$-clique is \emph{active} if it was never chosen before. At each time step, we select randomly an active $d+1$-clique and create a new vertex and connect this  new vertex to  every vertex of the $d+1$-clique chosen.  We repeat the procedure of selecting  active $d+1$-cliques and creating new vertices, until the network grows to a given size $N$. For the large $N$, the average degree of  HDRAN tends to $2(d+1)$.

The  HDRAN display the prominent properties observed in \spe{real-world} networks~\cite{ZhRoFr06}.  First, they are scale-free, since their degree distribution obeys  a power-law, with the  power exponent being $\gamma=2+\frac{1}{d-1}$. Second,  they are small-world, with their average path length growing logarithmically with the number of vertices. Third, their clustering coefficient is large and tends to a $d$-dependent constant,  increasing  with $d$. Finally, by construction, there are many cycles with various length in  HDRAN.

\begin{figure}
  \centering\includegraphics[width=\linewidth]{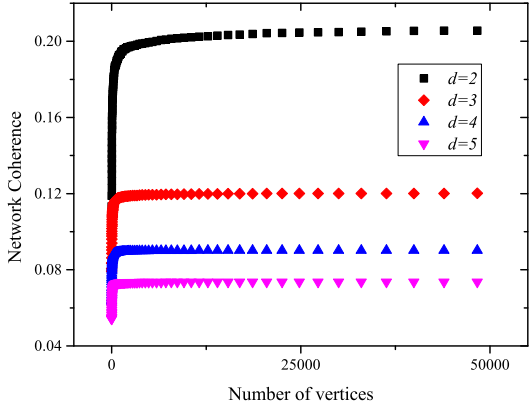}
  \caption{The coherence for  $d$-dimensional random Apollonian networks with various $d$.}\label{randSim.fig}
\end{figure}

In Figure~\ref{randSim.fig}, we report the numerical results  about  the coherence for HDRAN with  various $d$ and $N$.  From  Figure~\ref{randSim.fig}, we can observe that with the growth of  HDRAN,  the network coherence converges to a constant \spe{that depends} on the dimension parameter $d$: The higher the  dimension $d$, the lower the network coherence. This phenomenon is consistent with our intuition.

\section{Network Coherence for Deterministic Scale-free Networks}

To further uncover the behavior of coherence in  scale-free networks, in this section we derive closed form \spe{expressions} for the network coherence of two deterministic scale-free networks,  the pseudofractal scale-free web~\cite{DoGoMe02} and 4-clique motif scale-free network, both of which are constructed by edge iterations. These two networks are representative of a class of deterministic models for scale-free networks, since they display the classic properties observed in most real-\spe{world} networks.  Due to their special construction, we can derive exact results of coherence for both networks.


\subsection{Coherence for Pseudofractal Scale-free Web}

\begin{figure}
  \centering\includegraphics[width=7cm]{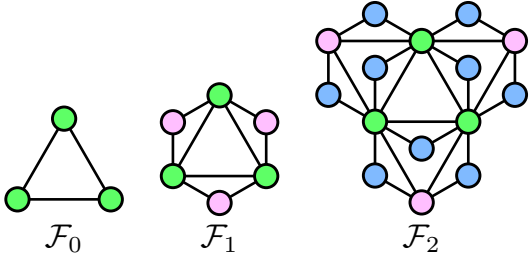}
  \caption{The first few iterations of  pseudofractal scale-free web.}\label{pseudo:fig}
\end{figure}

The  pseudofractal scale-free web after $g$ ($g \geq 0$) iterations, denoted by $\mathcal{F}_g$, is constructed in the following manner. Initially ($g = 0$), the network $\Fr_0$ consists of a triangle of  $3$ vertices and $3$ edges. At each iteration $g\geq 1$, for each existing edge $e$ in  $\mathcal{F}_{g-1}$ we add a new vertex and connect it to both end vertices of $e$. Figure \ref{pseudo:fig} shows networks $\Fr_0$, $\Fr_1$, and $\Fr_2$.  By construction, it is easy to verify that in  network $\Fr_g$, there are  $N_g=\frac{3^{g+1}+3}{2}$ vertices and $M_g=3^{g+1}$ edges. Then,  the average degree of $\Fr_g$ is $\rho=\frac{4 \times 3^{g+1}}{ 3^{g+1}+3}$, which is approximately $4$ when $g$ is  large enough.

The pseudofractal scale-free web is simultaneously scale-free and small-world~\cite{DoGoMe02}.  It has a power-law distribution with the power  exponent being $\gamma=1+\frac{\ln 3}{\ln 2}$. Its average path length grows logarithmically with the number of  vertices, when $g$ is large. Moreover, it has a large average clustering coefficient that  tends to a constant $\frac{4}{5}$ when $g$ is large.  In addition to these topological properties,
many combinatoric  properties of the pseudofractal scale-free web have also been well studied, such as  minimum dominating sets~\cite{ShLiZh17}, the number of spanning trees~\cite{ZhLiWuZh10}, and the distribution of cycles of different length~\cite{RoKiBoBe05}. 



\begin{theorem}
The network coherence for the pseudofractal scale-free web  $\Fr_g$,  $g\geq 0$, \spe{is}
\begin{align}
&H_{\mathrm{FO}}(\Fr_g)=\frac{1}{112\cdot 3^{g+2}(3^{g+1}+3)^2} \Big(50\cdot 3^{3g+3}  \nonumber\\
&\quad -35\cdot 3^{2g+2}2^{g+1}+48\cdot 3^{2g+2}+30\cdot 3^{g+2}2^{g+1}\nonumber\\
&\quad-14\cdot 3^{g+2}+225\cdot 2^{g+1}\Big)\,, \label{pseudoCoh}
\end{align}
which asymptotically converges to a constant when $g\to\infty$:
\begin{align}
\label{pseudoCohB}
\lim_{g\to\infty}H_{\mathrm{FO}}(\Fr_g)= \frac{25}{84}\,.
\end{align}
\end{theorem}
\begin{IEEEproof}
According to~\eqref{FOCohRes}, to find the coherence for  the pseudofractal scale-free web   $\Fr_g$, we can first determine its Kirchhoff index.  By replacing  $G$ in Theorem 5.3  in~\cite{YaKl15} with $\Fr_0$, we obtain the Kirchhoff index $R(\Fr_g)$ of $\Fr_g$, which reads
\begin{align}
\label{pseudoRes}
R(\Fr_g)=&\frac{1}{112} 3^{-(g+2)}\Big(50\cdot 3^{3g+3}-35\cdot 3^{2g+2}2^{g+1}\nonumber\\
&+48\cdot 3^{2g+2}+30\cdot 3^{g+2}2^{g+1}\nonumber\\
&\qquad\qquad -14\cdot 3^{g+2}+225\cdot 2^{g+1}\Big)\,.
\end{align}
Plugging~\eqref{pseudoRes} into~\eqref{FOCohRes}  leads to~\eqref{pseudoCoh}, which provides a closed form expression for $H_{\mathrm{FO}}(\Fr_g)$. When  $g\to\infty$, ~\eqref{pseudoCohB} is immediate from~\eqref{pseudoCoh}.
\end{IEEEproof}

We note that the lower bound for $H_{\mathrm{FO}}(\Fr_g)$, given in terms of  the average degree $\rho$, is $\frac{1}{2\rho}=\frac{1}{8}$.  Thus, the actual value $\frac{25}{84}$ of  the network coherence  $H_{\mathrm{FO}}(\Fr_g)$ is about as $2.38$ times  the  lower bound $\frac{1}{8}$, which means that  the pseudofractal scale-free web has an almost optimal structure  for noisy consensus without leaders.

\subsection{Coherence for 4-clique Motif  Scale-free Network}

In this subsection, we propose an variant of the  pseudofractal scale-free web and study its network coherence. The variant is also iteratively constructed and scale-free, which  consists of  4-vertex complete graphs, we thus call it  4-clique motif  scale-free network.
We next briefly introduce the construction method,  structural properties. and network coherence of first-order consensus dynamics for the  4-clique scale-free motif  network, while we omit the details of  computation or proofs due to the limitation of space. 

\begin{figure}
  \centering\includegraphics[width=7cm]{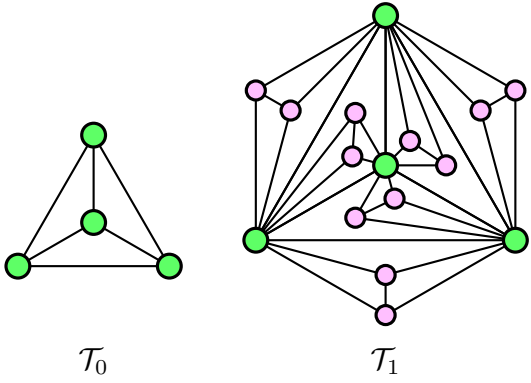}
  \caption{Illustration for the 4-clique motif networks  $\mathcal{T}_{0}$ and $\mathcal{T}_{1}$.}\label{example}
\end{figure}

We denote by $\mathcal{T}_g$ the 4-clique motif  scale-free network after $g$ ($g\geq 0$) iterations. Initially ($g=0$), $\mathcal{T}_0$ is a  4-vertex complete graph. For  $g\geq 1$,  $\mathcal{T}_g$ is obtained from $\mathcal{T}_{g-1}$ by performing the following operations:  for  each edge $e$ in $\mathcal{T}_{g-1}$, we create a new  edge $e'$ and connect each  end-vertex of  $e'$ to both  end-vertices of  edge  $e$.  For network $\mathcal{T}_g$,  let $V_g$ be the set of its vertices, let $\bar{V}_g$ be the set of vertices generated at  iteration $g$, and let $E_g$ be the set of  all edges. It is easy to derive that in network  $\mathcal{T}_g$ there are  $N_{g}=|V_g|=\frac{2}{5}\left(6^{g+1}+4\right)$ vertices and $M_{g}= 6^{g+1}$ edges. Thus the average degree of $\mathcal{T}_g$ is $\rho=\frac{5\times 6^{g+1}}{6^{g+1}+4}$, which  \spe{is approximately}  $5$ when $g$ is  large. Figure \ref{example} illustrates  networks $\mathcal{T}_{0}$ and $\mathcal{T}_{1}$.



The 4-clique motif  scale-free network also \spe{displays} some common properties \spe{of real-world networks}. It has a power-law degree distribution with  exponent $\gamma$ equal to $2+\frac{\ln 2}{\ln 3}$. It \spe{has} the  small-world effect, with the average path length growing as a logarithmic function of the number of vertices and  the clustering coefficient tending to a high value $ \frac{265}{306}$. Thus, both networks $\mathcal{T}_{g}$ and $\mathcal{F}_{g}$ exhibit similar structural properties; it \spe{is} then expected that their dynamical properties (e.g., network coherence) also bear a strong resemblance to each other.

We now derive the expression for the \spe{coherence} $H_{\mathrm{FO}}$ of  first-order noisy consensus  in $\mathcal{T}_g$. We will  show that $H_{\mathrm{FO}}(\mathcal{T}_g)$ converges to a constant with the growth of the network. To this end, we first determine the Kirchhoff index for $\mathcal{T}_g$.  Let $\Omega_{ij}^{(g)}$ be the effective resistance between a pair of nodes $i$ and $j$ in $\mathcal{T}_g$. For network $\mathcal{T}_{g+1}$,  the following Lemma  gives the evolution of effective resistance between any pair of old nodes, namely, those nodes already present  in $\mathcal{T}_{g}$, and shows that the effective resistance between any two nodes in $\mathcal{T}_{g+1}$ can be expressed in terms of those  pairs of old nodes in $\mathcal{T}_{g}$.

\begin{lemma}
\label{pairRes}
Let $\{i,j\}$ be a unordered pair of vertices in $\mathcal{T}_{g+1}$; the effective resistance $\Omega_{ij}^{(g+1)}$ satisfies:
\begin{enumerate}
\item If $i,j\in V_g$, then $\Omega_{ij}^{(g+1)}=\frac{1}{2}\Omega_{ij}^{(g)}$\,.
\item If $i,j\in \bar{V}_{g+1}$ and $\{i,j\}\in E_{g+1}$, then $\Omega^{(g+1)}_{ij}=\frac{1}{2}$.
\item If $i\in \bar{V}_{g+1}$, $j,l\in {\mathcal{N}}_i \cap V_g$, then $\Omega^{(g+1)}_{ij}=\frac{\Omega^{(g)}_{jl}}{8}+\frac{3}{8}$.
\item If $i\in \bar{V}_{g+1}$, $j \in V_g$, $k,l\in  {\mathcal{N}}_i \cap V_g$, then
\begin{equation*}
\Omega^{(g+1)}_{ij}=\frac{3+2\Omega^{(g)}_{jl}+2\Omega^{(g)}_{jk}-\Omega^{(g)}_{kl}}{8}\,,
\end{equation*}
which is also true for $j\in \{k,l\}$, and  is reduced to case 3).
\item If $i,j\in \bar{V}_{g+1}$, $\{i,j\}\notin E_{g+1}$,  $k,l\in {\mathcal{N}}_i \cap V_g$, and $p,q \in {\mathcal{N}}_j \cap V_g$, then
\begin{small}
\begin{align*}
\Omega^{(g+1)}_{ij}=&\frac{6+\Omega^{(g)}_{kq}+\Omega^{(g)}_{kp}+\Omega^{(g)}_{lp}+\Omega^{(g)}_{lq}-(\Omega^{(g)}_{kl} + \Omega^{(g)}_{pq})}{8}\,.
\end{align*}
\end{small}
\end{enumerate}
\end{lemma}

Using Lemmas~\ref{Foster},~\ref{sumRule}, and~\ref{pairRes}, we can derive the recursive relations for the Kirchhoff index $R(\mathcal{T}_{g})$,   multiplicative degree-Kirchhoff index $R^{\ast}(\mathcal{T}_{g})$, and additive degree-Kirchhoff index $R^{+}(\mathcal{T}_{g})$ for  network $\mathcal{T}_g$, and thus obtain their exact expressions in terms of $g$.

\begin{lemma}
\label{KirRes.lemma}
The multiplicative degree-Kirchhoff index, additive degree-Kirchhoff index, and Kirchhoff index of network  $\mathcal{T}_{g}$ are, respectively,
\begin{small}
\begin{align}
\label{mulKirRes}
R^{\ast}(\mathcal{T}_{g})=&\frac{1}{5}\cdot 2^g 3^{g+2} \left(13\cdot 2^{g+1} 3^g-5\cdot  3^{g+1}+4\right),\\
\label{addKirRes}
R^{+}(\mathcal{T}_{g})=&\frac{9}{275} \Big(169\cdot 2^{2 g+2} 3^{2 g} -55\cdot 2^{g+1} 3^{2g+1}\nonumber\\
&+11\cdot 2^{g+3} 3^g+35\cdot 3^{g+1}+11\Big),\\
\label{kirRes}
R(\mathcal{T}_{g})=&\frac{3}{275} \Big(13\cdot 2^{2 g+1} 3^{2 g+2}-11\cdot 2^{g} 3^{2 g+2}\nonumber\\
\quad\qquad &+13\cdot 2^{g+2} 3^g+7\cdot 3^{g+2}-11+36\cdot 2^{-g}\Big).
\end{align}
\end{small}
\end{lemma}
\begin{IEEEproof}
Using the technique and procedure similar to those in~\cite{YaKl15}, we can derive  the following recursive relations governing the evolution for $R(\mathcal{T}_{g})$, $R^{\ast}(\mathcal{T}_{g})$, and $R^{+}(\mathcal{T}_{g})$.
\begin{align}
\label{Res}
R(\mathcal{T}_{g+1})
=&\frac{3}{2}M_{g}^2-\frac{1}{4}N_{g}(N_{g}-1)+\frac{1}{4} M_{g}(N_{g}-2)\nonumber\\
&+\frac{1}{2}\left(R(\mathcal{T}_{g})+R^{+}(\mathcal{T}_{g})+R^{\ast}(\mathcal{T}_{g})\right)\,,\\
\label{Rmu}
R^{\ast}(\mathcal{T}_{g+1})=&9\left(3M_{g}^{2}-M_{g}N_{g}\right)+18 R^{\ast}(\mathcal{T}_{g})\,,\\
\label{Rad}
R^{+}(\mathcal{T}_{g+1})
=&\frac{27}{2}M_{g}^2-\frac{3}{2}M_{g}-\frac{3}{4}N_{g}(N_{g}-1)-\frac{9}{4}M_{g}N_{g}\nonumber\\
& +(3R^{+}(\mathcal{T}_{g})+6R^{\ast}(\mathcal{T}_{g}))\,.
\end{align}
Considering  $N_{g}=\frac{2}{5}\left(6^{g+1}+4\right)$, $M_{g}= 6^{g+1}$,  and the initial conditions $R^{\ast}(\mathcal{T}_{0})=27$, $R^{+}(\mathcal{T}_{0})=18$, and $R(\mathcal{T}_{0})=3$, the above recursive equations~\eqref{Res},~\eqref{Rmu}, and~\eqref{Rad} are solved to obtain the expressions for $R^{\ast}(\mathcal{T}_{g})$, $R^{+}(\mathcal{T}_{g})$, and $R(\mathcal{T}_{g})$ as given   in the lemma.
\end{IEEEproof}

Lemma~\ref{KirRes.lemma}, together with~\eqref{FOCohRes}, \spe{leads} to the following theorem.
\begin{theorem}
For network $\mathcal{T}_g$, $g\geq 0$, the first-order coherence   is
\begin{align}
H_{\mathrm{FO}}(\mathcal{T}_g)=&\frac{3}{88}(6^{g+1}+4)^{-2} \Big(13\cdot 2^{2 g+1} 3^{2 g+2}-11\cdot 2^{g} 3^{2 g+2}\nonumber\\
&+13\cdot 2^{g+2} 3^g+7\cdot 3^{g+2}-11+36\cdot 2^{-g}\Big).
\end{align}
In the limit of large $g$ ($g\to\infty$),
\begin{align}
\lim_{g\to \infty}H_{\mathrm{FO}}(\mathcal{T}_g) =\frac{39}{176}\,.
\end{align}
\end{theorem}

Therefore, in large networks $\mathcal{T}_g$, the first-order network coherence converges to a small constant $\frac{39}{176}$, much smaller than 1.
We remark that  the lower bound for $H_{\mathrm{FO}}(\mathcal{T}_g)$ given in terms of  the average degree  is $\frac{1}{10}$, lower than the  actual value $\frac{39}{176}$, with the   actual value  being about  $2.22$ times  the  lower bound. Thus, similar to the   pseudofractal scale-free web, the 4-clique motif scale-free networks also has an almost optimal structure  for noisy consensus dynamics.

\section{Result Analysis}\label{analysis.sec}

In the preceding sections, we have studied the first-order noisy consensus dynamics in \spe{real-world} and  model scale-free networks. The results show that in all considered networks, the network coherence is very small and does not depend on the number of vertices, but converges to constants. Thus, all \spe{of} these networks are almost optimal in the sense that they are very robust to Gaussian noisy in  consensus dynamics. Since the  first-order coherence of a network is determined by all non-zero eigenvalues of its Laplacian matrix, which is in turn influenced by the structure of the network, the small constant coherence observed for studied networks are due to their structural properties, especially the scale-free behavior and cycles of various length,  which can be understood from the following heuristic arguments.

As shown in~\eqref{FOCohRes},  the  coherence is closely related to resistance distance, which is bounded by the shortest-path distance. In a  scale-free network, there exist  vertices with large degree that are linked to many other vertices in the network, which results in the small-world phenomenon that  the average path length grows at most  logarithmically with  the number of vertices in the network~\cite{Ne03}.  Moreover, in many realistic and model scale-free networks, there are many cycles  with different lengths, which lead to various alternative paths of different lengths between many vertex pairs. As a result, the average resistance distance over all pairs of vertices does not increase with the growth of the network. Below, by way of illustration we  show that neither power-law behavior nor  cycles alone can guarantee  a constant network coherence.

In~\cite{YiZhLiCh15}, it is shown that in the small-world Farey network, the network coherence  $H_{\mathrm{FO}}$ scales logarithmically with the number of vertices $N$ as $ H_{\mathrm{FO}} \sim \ln N$. However, in the pseudofractal scale-free web,  $H_{\mathrm{FO}}$ is a constant, independent of $N$. Note that both Farey network and pseudofractal scale-free web are small-world and highly clustered. Moreover, there are many cycles at different scales in both networks. The  reason for the distinction of their coherence lies in, at least partially, the scale-free property of the
pseudofractal scale-free web that is absent in the Farey network. On the other hand, the coherence $H_{\mathrm{FO}}$ for the Koch network~\cite{YiZhShCh17} is a logarithmic function of $N$, despite its scale-free structure.  The reason the  coherence for the Koch network is not a constant is  because it has only triangles,  lacking cycles of different length.

\section{Conclusion}\label{conclusion.sec}

Various real-life networks are sparse and loopy, displaying the striking scale-free and small-world properties, which heavily affect the behaviors of diverse dynamics occurring on these networks. In this paper, we have studied  first-order consensus dynamics with disturbances in scale-free networks, with an emphasis on the network coherence. We first provided a lower bound and an upper bound for coherence in a general network, with the lower bound being half of the reciprocal of the average degree. We then studied the network coherence for some representative real-world networks with the scale-free property, and found that their coherence is very small, irrespective of the size of the networks. We also studied numerically the coherence for Barab\'{a}si-Albert networks and high dimensional Apollonian networks, which converges to constants when the networks grow. Finally, we studied analytically the coherence for two deterministic scale-free networks, obtaining  explicit expressions for the network coherence, which converge to small constants. Our results indicate that scale-free networks are resistant to stochastic disturbances imposed on the consensus algorithm. We argued that the scale-free and loopy structure is responsible for the robustness against the noise.

\appendices

\bibliographystyle{IEEEtran}
\bibliography{consensus}

\begin{thebibliography}{10}
\providecommand{\url}[1]{#1}
\csname url@rmstyle\endcsname
\providecommand{\newblock}{\relax}
\providecommand{\bibinfo}[2]{#2}
\providecommand\BIBentrySTDinterwordspacing{\spaceskip=0pt\relax}
\providecommand\BIBentryALTinterwordstretchfactor{4}
\providecommand\BIBentryALTinterwordspacing{\spaceskip=\fontdimen2\font plus
\BIBentryALTinterwordstretchfactor\fontdimen3\font minus
  \fontdimen4\font\relax}
\providecommand\BIBforeignlanguage[2]{{%
\expandafter\ifx\csname l@#1\endcsname\relax
\typeout{** WARNING: IEEEtran.bst: No hyphenation pattern has been}%
\typeout{** loaded for the language `#1'. Using the pattern for}%
\typeout{** the default language instead.}%
\else
\language=\csname l@#1\endcsname
\fi
#2}}

\bibitem{Ly97}
N.~A. Lynch, \emph{Distributed algorithms}.\hskip 1em plus 0.5em minus
  0.4em\relax San Francisco, CA: Morgan Kaufmann, 1996.

\bibitem{Ts84}
J.~N. Tsitsiklis, ``Problems in decentralized decision making and
  computation,'' Ph.D. dissertation, Massachusetts Institute of Technology,
  1984.

\bibitem{SaFaMu07}
R.~Olfati-Saber, J.~A. Fax, and R.~M. Murray, ``Consensus and cooperation in
  networked multi-agent systems,'' \emph{Proc. IEEE}, vol.~95, no.~1, pp.
  215--233, Jan. 2007.

\bibitem{SuStErBrGh15}
W.~Sun, E.~G. Str{\"o}m, F.~Br{\"a}nnstr{\"o}m, and M.~R. Gholami, ``Random
  broadcast based distributed consensus clock synchronization for mobile
  networks,'' \emph{IEEE Trans. Wirel. Commun.}, vol.~14, no.~6, pp.
  3378--3389, 2015.

\bibitem{Cy89}
G.~Cybenko, ``Dynamic load balancing for distributed memory multiprocessors,''
  \emph{J. Parallel Distrib. Comput.}, vol.~7, no.~2, pp. 279--301, 1989.

\bibitem{MuGhSc98}
S.~Muthukrishnan, B.~Ghosh, and M.~H. Schultz, ``First-and second-order
  diffusive methods for rapid, coarse, distributed load balancing,''
  \emph{Theor. Comput. Syst.}, vol.~31, no.~4, pp. 331--354, 1998.

\bibitem{DiFrMo99}
R.~Diekmann, A.~Frommer, and B.~Monien, ``Efficient schemes for nearest
  neighbor load balancing,'' \emph{Parallel Comput.}, vol.~25, no.~7, pp.
  789--812, 1999.

\bibitem{AmFrJiVe15}
N.~Amelina, A.~Fradkov, Y.~Jiang, and D.~J. Vergados, ``Approximate consensus
  in stochastic networks with application to load balancing,'' \emph{IEEE
  Trans. Inf. Theory}, vol.~61, no.~4, pp. 1739--1752, 2015.

\bibitem{FaMu04}
J.~A. Fax and R.~M. Murray, ``Information flow and cooperative control of
  vehicle formations,'' \emph{IEEE Trans. Autom. Control}, vol.~49, no.~9, pp.
  1465--1476, Sep. 2004.

\bibitem{Sa06}
R.~Olfati-Saber, ``Flocking for multi-agent dynamic systems: Algorithms and
  theory,'' \emph{IEEE Trans. Autom. Control}, vol.~51, no.~3, pp. 401--420,
  Mar. 2006.

\bibitem{DiKy07}
D.~V. Dimarogonas and K.~J. Kyriakopoulos, ``On the rendezvous problem for
  multiple nonholonomic agents,'' \emph{IEEE Trans. Autom. Control}, vol.~52,
  no.~5, pp. 916--922, May 2007.

\bibitem{GiPa16}
L.~F. Giraldo and K.~M. Passino, ``Dynamic task performance, cohesion, and
  communications in human groups,'' \emph{IEEE Trans. Cybern.}, vol.~46,
  no.~10, pp. 2207--2219, 2016.

\bibitem{LiRu06}
Q.~Li and D.~Rus, ``Global clock synchronization in sensor networks,''
  \emph{IEEE Trans. Comput.}, vol.~55, no.~2, pp. 214--226, 2006.

\bibitem{YuChWaYa09}
W.~Yu, G.~Chen, Z.~Wang, and W.~Yang, ``Distributed consensus filtering in
  sensor networks,'' \emph{IEEE Trans. Syst., Man, Cybern. B, Cybern.},
  vol.~39, no.~6, pp. 1568--1577, 2009.

\bibitem{ZhChLiYaGu13}
S.~Zhu, C.~Chen, W.~Li, B.~Yang, and X.~Guan, ``Distributed optimal consensus
  filter for target tracking in heterogeneous sensor networks,'' \emph{IEEE
  Trans. Cybern.}, vol.~43, no.~6, pp. 1963--1976, 2013.

\bibitem{AL14}
A.~H. Sayed, ``Adaptation, learning, and optimization over networks,''
  \emph{Foundations and Trends in Machine Learning}, vol.~7, no. 4-5, pp.
  311--801, 2014.

\bibitem{MaBrMaSa16}
V.~Matta, P.~Braca, S.~Marano, and A.~H. Sayed, ``Diffusion-based adaptive
  distributed detection: Steady-state performance in the slow adaptation
  regime,'' \emph{IEEE Trans. Inf. Theory}, vol.~62, no.~8, pp. 4710--4732,
  2016.

\bibitem{BiBlCa16}
G.~Biau, K.~Bleakley, and B.~Cadre, ``The statistical performance of
  collaborative inference,'' \emph{J. Mach. Learn. Res.}, vol.~17, no.~1, pp.
  2200--2228, Jan. 2016.

\bibitem{MoTa14}
S.~Motsch and E.~Tadmor, ``Heterophilious dynamics enhances consensus,''
  \emph{SIAM Rev.}, vol.~56, no.~4, pp. 577--621, 2014.

\bibitem{WuTaCaZh16}
X.~Wu, Y.~Tang, J.~Cao, and W.~Zhang, ``Distributed consensus of stochastic
  delayed multi-agent systems under asynchronous switching,'' \emph{IEEE Trans.
  Cybern.}, vol.~46, no.~8, pp. 1817--1827, 2016.

\bibitem{PaBa10}
S.~Patterson and B.~Bamieh, ``Leader selection for optimal network coherence,''
  in \emph{Proc. 49th {IEEE} Conf. Decision Control}.\hskip 1em plus 0.5em
  minus 0.4em\relax IEEE, 2010, pp. 2692--2697.

\bibitem{PaBa11}
------, ``Network coherence in fractal graphs,'' in \emph{Proc. 50th {IEEE}
  Conf. Decision Control}, Dec. 2011, pp. 6445--6450.

\bibitem{BaJoMiPa12}
B.~Bamieh, M.~Jovanovic~R, P.~Mitra, and S.~Patterson, ``Coherence in
  large-scale networks: Dimension-dependent limitations of local feedback,''
  \emph{{IEEE} Trans. Autom. Control}, vol.~57, no.~9, pp. 2235--2249, Sep.
  2012.

\bibitem{PaBa14}
S.~Patterson and B.~Bamieh, ``Consensus and coherence in fractal networks,''
  \emph{IEEE Trans. Control Netw. Syst.}, vol.~1, no.~4, pp. 338--348, Sep.
  2014.

\bibitem{XiBoKi07}
L.~Xiao, S.~Boyd, and S.-J. Kim, ``Distributed average consensus with
  least-mean-square deviation,'' \emph{J. Parallel. Distrib. Comput.}, vol.~67,
  no.~1, pp. 33--46, Jan. 2007.

\bibitem{BaJoMiPa08}
B.~Bamieh, M.~Jovanovic~R, P.~Mitra, and S.~Patterson, ``Effect of topological
  dimension on rigidity of vehicle formations: Fundamental limitations of local
  feedback,'' in \emph{Proc. 47th {IEEE} Conf. Decision Control}, Dec. 2008,
  pp. 369--374.

\bibitem{YoScLe10}
G.~F. Young, L.~Scardovi, and N.~E. Leonard, ``Robustness of noisy consensus
  dynamics with directed communication,'' in \emph{Proc. Amer. Control Conf.},
  Jun. 2010, pp. 6312--6317.

\bibitem{RaWa11}
R.~Rajagopal and M.~J. Wainwright, ``Network-based consensus averaging with
  general noisy channels,'' \emph{IEEE Trans. Signal Process.}, vol.~59, no.~1,
  pp. 373--385, 2011.

\bibitem{YaBl13}
Y.~Yang and R.~S. Blum, ``Broadcast-based consensus with non-zero-mean
  stochastic perturbations,'' \emph{IEEE Trans. Inf. Theory}, vol.~59, no.~6,
  pp. 3971--3989, 2013.

\bibitem{YiZhLiCh15}
Y.~Yi, Z.~Zhang, Y.~Lin, and G.~Chen, ``Small-world topology can significantly
  improve the performance of noisy consensus in a complex network,''
  \emph{Comput. J.}, vol.~58, no.~12, pp. 3242--3254, 2015.

\bibitem{HeZhChShCh16}
J.~He, M.~Zhou, P.~Cheng, L.~Shi, and J.~Chen, ``{Consensus under bounded noise
  in discrete network systems: An algorithm with fast convergence and high
  accuracy},'' \emph{IEEE Trans. Cybern.}, vol.~46, no.~12, pp. 2874--2884,
  2016.

\bibitem{YiZhShCh17}
Y.~Yi, Z.~Zhang, L.~Shan, and G.~Chen, ``Robustness of first-and second-order
  consensus algorithms for a noisy scale-free small-world {K}och network,''
  \emph{IEEE Trans. Control Syst. Technol.}, vol.~25, no.~1, pp. 342--350,
  2017.

\bibitem{ZhCo11}
Z.~Zhang and F.~Comellas, ``Farey graphs as models for complex networks,''
  \emph{Theor. Comput. Sci.}, vol. 412, no.~8, pp. 865--875, Mar. 2011.

\bibitem{Ne03}
M.~E.~J. Newman, ``{The structure and function of complex networks},''
  \emph{{SIAM Rev.}}, vol.~{45}, no.~{2}, pp. {167--256}, {Jun.} {2003}.

\bibitem{BaAl99}
A.-L. Barab{\'a}si and R.~Albert, ``Emergence of scaling in random networks,''
  \emph{Science}, vol. 286, no. 5439, pp. 509--512, 1999.

\bibitem{WaSt98}
D.~J. Watts and S.~H. Strogatz, ``Collective dynamics of `small-world'
  networks,'' \emph{Nature}, vol. 393, no. 6684, pp. 440--442, {Jun.} 1998.

\bibitem{RoKiBoBe05}
H.~D. Rozenfeld, J.~E. Kirk, E.~M. Bollt, and D.~Ben-Avraham, ``Statistics of
  cycles: how loopy is your network?'' \emph{J. Phys. A}, vol.~38, no.~21, p.
  4589, 2005.

\bibitem{KlSt06}
K.~Klemm and P.~F. Stadler, ``Statistics of cycles in large networks,''
  \emph{Phys. Rev. E}, vol.~73, no.~2, p. 025101, 2006.

\bibitem{ZhWu15}
Z.~Zhang and B.~Wu, ``Pfaffian orientations and perfect matchings of scale-free
  networks,'' \emph{Theor. Comput. Sci.}, vol. 570, pp. 55--69, 2015.

\bibitem{JiLiZh17}
Y.~Jin, H.~Li, and Z.~Zhang, ``{Maximum matchings and minimum dominating sets
  in Apollonian networks and extended Tower of Hanoi graphs},'' \emph{Theoret.
  Comput. Sci.}, vol. 703, pp. 37--54, 2017.

\bibitem{PaVe01}
R.~Pastor-Satorras and A.~Vespignani, ``Epidemic spreading in scale-free
  networks,'' \emph{Phys. Rev. Lett.}, vol.~86, no.~14, p. 3200, 2001.

\bibitem{SaSaPa08}
F.~C. Santos, M.~D. Santos, and J.~M. Pacheco, ``Social diversity promotes the
  emergence of cooperation in public goods games,'' \emph{Nature}, vol. 454,
  no. 7201, pp. 213--216, 2008.

\bibitem{LiSlBa11}
Y.-Y. Liu, J.-J. Slotine, and A.-L. Barab{\'a}si, ``Controllability of complex
  networks,'' \emph{Nature}, vol. 473, no. 7346, pp. 167--173, 2011.

\bibitem{ZhYaWa05}
T.~Zhou, G.~Yan, and B.-H. Wang, ``Maximal planar networks with large
  clustering coefficient and power-law degree distribution,'' \emph{Phys. Rev.
  E}, vol.~71, no.~4, p. 046141, 2005.

\bibitem{ZhRoFr06}
Z.~Zhang, L.~Rong, and F.~Comellas, ``High-dimensional random {A}pollonian
  networks,'' \emph{Physica A}, vol. 364, pp. 610--618, 2006.

\bibitem{DoGoMe02}
S.~N. Dorogovtsev, A.~V. Goltsev, and J.~F.~F. Mendes, ``Pseudofractal
  scale-free web,'' \emph{Phys. Rev. E}, vol.~65, no.~6, p. 066122, 2002.

\bibitem{Me98}
R.~Merris, ``Laplacian graph eigenvectors,'' \emph{Linear Algebra Appl.}, vol.
  278, no. 1-3, pp. 221--236, 1998.

\bibitem{KlRa93}
D.~J. Klein and M.~Randi{\'c}, ``Resistance distance,'' \emph{J. Math. Chem.},
  vol.~12, no.~1, pp. 81--95, 1993.

\bibitem{Fo49}
R.~M. Foster, ``The average impedance of an electrical network,''
  \emph{Contributions to Applied Mechanics (Reissner Anniversary Volume)}, pp.
  333--340, 1949.

\bibitem{Ch10}
H.~Chen, ``Random walks and the effective resistance sum rules,''
  \emph{Discrete Appl. Math.}, vol. 158, no.~15, pp. 1691--1700, 2010.

\bibitem{ChZh07}
H.~Chen and F.~Zhang, ``{Resistance distance and the normalized Laplacian
  spectrum},'' \emph{Discrete Appl. Math.}, vol. 155, no.~5, pp. 654--661,
  2007.

\bibitem{GuFeYu12}
I.~Gutman, L.~Feng, and G.~Yu, ``Degree resistance distance of unicyclic
  graphs,'' \emph{Trans. Combin.}, vol.~1, no.~2, pp. 27--40, 2012.

\bibitem{LoZa10}
E.~Lovisari and S.~Zampieri, ``Performance metrics in the consensus problem: a
  survey,'' \emph{IFAC Proceedings Volumes}, vol.~43, no.~21, pp. 324--335,
  2010.

\bibitem{Me94}
R.~Merris, ``Laplacian matrices of graphs: a survey,'' \emph{Linear Algebra
  Appl.}, vol. 197, pp. 143--176, 1994.

\bibitem{Ku13}
J.~Kunegis, ``Konect: The koblenz network collection,'' in \emph{Proc. 22nd
  Int. Conf. World Wide Web}, 2013, pp. 1343--1350.

\bibitem{BiCa03}
G.~Bianconi and A.~Capocci, ``Number of loops of size $h$ in growing scale-free
  networks,'' \emph{Phys. Rev. Lett.}, vol.~90, no.~7, p. 078701, 2003.

\bibitem{ShLiZh17}
L.~Shan, H.~Li, and Z.~Zhang, ``Domination number and minimum dominating sets
  in pseudofractal scale-free web and {S}ierpi{\'n}ski graph,'' \emph{Theoret.
  Comput. Sci.}, vol. 677, pp. 12--30, 2017.

\bibitem{ZhLiWuZh10}
Z.~Zhang, H.~Liu, B.~Wu, and S.~Zhou, ``Enumeration of spanning trees in a
  pseudofractal scale-free web,'' \emph{EPL (Europhysics Letters)}, vol.~90,
  no.~6, p. 68002, 2010.

\bibitem{YaKl15}
Y.~Yang and D.~J. Klein, ``Resistance distance-based graph invariants of
  subdivisions and triangulations of graphs,'' \emph{Discrete Appl. Math.},
  vol. 181, pp. 260--274, 2015.

\end{thebibliography}

\end{document}